\documentclass[twocolumn,amssymb,amsmath,aps,secnumarabic,floatfix]{revtex4}
\usepackage{psfig}
\usepackage{epsfig}
\usepackage{pstricks}
\usepackage{dcolumn}
\newpsobject{showgrid}{psgrid}{subgriddiv=1}
\expandafter\ifx\csname package@font\endcsname\relax\else
 \expandafter\expandafter
 \expandafter\usepackage
 \expandafter\expandafter
 \expandafter{\csname package@font\endcsname}
\fi
\begin{document}
\title{Comment on: ``Superscaling of Percolation on Rectangular Domains''}
\author{Gunnar Pruessner}
\email{gunnar.pruessner@physics.org}
\affiliation{
Virginia Polytechnic Inst. \& State Univ.,
Physics Department - Robeson Hall,
Blacksburg, VA 24061-0435,
USA}
\author{Nicholas R. Moloney}
\email{moloney@general.elte.hu}
\affiliation{
Institute for Theoretical Physics,
E\"{o}tv\"{o}s University,
P\'{a}zm\'{a}ny P\'{e}ter s\'{e}t\'{a}ny 1/a,
1117 Budapest,
Hungary,
EU
}

\date{\today}
\maketitle

In \cite{WatanabeETAL:2004}, the authors show numerically that spanning and
percolation probabilities in two-dimensional systems with different aspect
ratios obey a form of ``superscaling''. This scaling form makes it possible
to relate the percolation properties of a system with one set of parameters
$(L,\epsilon,R)$, to a system with another set of parameters.

In this comment, we would like to point out some difficulties with their
proposed scaling ansatz and suggest why this remained undetected in their
numerical analysis. Starting from the central result for the existence
probability Eq.~(8),
one observes that it
cannot account for Cardy's exact result \cite{Cardy:1992}, because it states
that $E_p$ is independent of the aspect ratio $R$ 
at the critical point, where $\epsilon=0$. 

Moreover, a simple analysis shows that the proposed scaling form for
$E_p$ does not behave as expected in the limit of large $R$. This can
be seen as follows: The relation $E_p(L,\epsilon,R) = E_p(L
u^{-a/y_t}, \epsilon, R u)$ follows directly from the scaling
assumption 
for $u>1$.
The left
hand side describes the probability of finding a cluster that spans
any two opposite sides of a system of size $L R \times L$ at bond
density $\rho=\epsilon+\rho_c$, while the right hand side describes
the same probability for a system of size $L R u^{1-a/y_t} \times L
u^{-a/y_t}$. Choosing $\rho$ small enough such that
$E_p(L,\epsilon,R)$ is very small, it is clear that it must be
exceeded at some point by $E_p(L u^{-a/y_t}, \epsilon, R u)$ in the
limit of large $u$.

\begin{figure}[h!]
\includegraphics[width=0.8\linewidth]{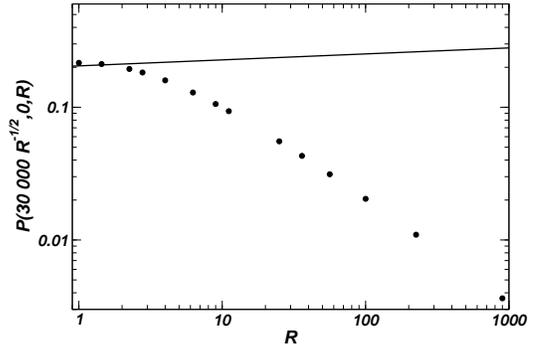}
\caption{
Percolation probability at $\rho=\rho_c$ for bond percolation, system size
$30\,000 R^{1/2}\times30\,000 R^{-1/2}$. The full line shows a comparison to Eq.~(10) in
\cite{WatanabeETAL:2004}, $P(30\,000 R^{-1/2},0,R)\propto
R^{-\beta(-y_t/2+b)}$ with $y_t=3/4$, $\beta=5/36$ and $b=0.05$.
\label{fig:our_measurements}
}
\end{figure}

Similar problems apply to the percolation probability $P$, discussed around
Eq. (10) in \cite{WatanabeETAL:2004}. Using our earlier results
\cite{MoloneyPruessner:2003}, we could verify that the scaling ansatz (10) does
not apply at $\epsilon=0$, see Fig.~\ref{fig:our_measurements}. 
Also, while 
$P(L,\epsilon,R) = P(LR, \epsilon, R^{-1})$, this gives for the scaling function $F$ in
$P(L,\epsilon,R) \propto (L^{y_t} R^b)^{-x} F ( \epsilon L^{y_t} R^b)$
that $F(1)=R^{(2b-y_t)x} F(R^{y_t-2b})$, which leads to a non-trivial scaling
function only if $b=y_t/2=3/8$, far from $b=0.05$
\cite{WatanabeETAL:2004}.

We suggest that these inconsistencies did not show up in the numerical
analysis because of the way in which the authors ``correct for
finite-size effects''. After introducing this correction in Eq. (9),
their data analysis is based on an $\epsilon=\rho-\rho_c$ in the scaling
function
shifted by a small amount
to $\epsilon'=\rho-\rho_c'(L,R)$ where $\rho_c'(L,R)$ is defined implicitly by
  $E_p(L,\rho_c'(L,R)-\rho_c,R)=E_p(L,0,1)$.
However, this shift is \emph{not a finite size correction} because,
\emph{regardless of the size}, it is a necessary adjustment in
order to produce the intended superscaling behaviour.

Although the \emph{shift vanishes in the thermodynamic limit},
$\lim_{L\to\infty} \rho_c'-\rho_c=0$, it remains
crucial; the \emph{difference} between the values of the observable with and
without shift \emph{does not vanish in the thermodynamic limit}, but converges
to a finite constant, for example $\lim_{L\to\infty}
E_p(L,\rho_c'-\rho_c,R)-E_p(L,0,R)\ne0$ if $R\ne1$. This is possible because
$\lim_{L\to\infty}E_p(L,\rho-\rho_c,R)$ is discontinuous at $\rho-\rho_c=0$. 

Allowing for such a shift undermines the notion
of universality: In fact, there is a $\rho_c'(L,R)$ that deviates from $\rho_c$
only by $\mathcal{O}(L^{-y_t})$ such that $E_p(L,\rho_c'-\rho_c,R)$ or
$P(L,\rho_c'-\rho_c,R)$ is equal to any arbitrary constant $0<c<1$.

What the authors actually show is that, for example,
$E_p(L,\epsilon,R) \propto F( ( \epsilon + \Delta \epsilon (L, R))
L^{y_t} R^a)$
where $\Delta \epsilon = \rho_c' - \rho_c$, which is in fact a function of
system size and aspect ratio.  
This limitation of their result becomes even
clearer in the other scaling form they suggest which takes ``into account
higher order corrections to the scaling'',
$E_p(L,\epsilon,R) \propto F(c_0(L,R) + \rho c_1(L,R) + \rho^2 c_2(L,R)
)$.
This form, while interesting in itself, is not a form of
scaling and therefore should not be called ``superscaling''.

GP would like to thank the
Alexander von Humboldt foundation as well as the NSF
(DMR-0308548/0414122) for support.
\bibliography{articles,books}

\begin{thebibliography}{3}
\expandafter\ifx\csname natexlab\endcsname\relax\def\natexlab#1{#1}\fi
\expandafter\ifx\csname bibnamefont\endcsname\relax
  \def\bibnamefont#1{#1}\fi
\expandafter\ifx\csname bibfnamefont\endcsname\relax
  \def\bibfnamefont#1{#1}\fi
\expandafter\ifx\csname citenamefont\endcsname\relax
  \def\citenamefont#1{#1}\fi
\expandafter\ifx\csname url\endcsname\relax
  \def\url#1{\texttt{#1}}\fi
\expandafter\ifx\csname urlprefix\endcsname\relax\def\urlprefix{URL }\fi
\providecommand{\bibinfo}[2]{#2}
\providecommand{\eprint}[2][]{\url{#2}}

\bibitem[{\citenamefont{Watanabe et~al.}(2004)\citenamefont{Watanabe, Yukawa,
  Ito, and Hu}}]{WatanabeETAL:2004}
\bibinfo{author}{\bibfnamefont{H.}~\bibnamefont{Watanabe}},
  \bibinfo{author}{\bibfnamefont{S.}~\bibnamefont{Yukawa}},
  \bibinfo{author}{\bibfnamefont{N.}~\bibnamefont{Ito}}, \bibnamefont{and}
  \bibinfo{author}{\bibfnamefont{C.-K.} \bibnamefont{Hu}},
  \bibinfo{journal}{Phys.~Rev.~Lett.} \textbf{\bibinfo{volume}{93}},
  \bibinfo{pages}{190601} (\bibinfo{year}{2004}).

\bibitem[{\citenamefont{Cardy}(1992)}]{Cardy:1992}
\bibinfo{author}{\bibfnamefont{J.}~\bibnamefont{Cardy}},
  \bibinfo{journal}{J.~Phys.~A:~Math.~Gen.} \textbf{\bibinfo{volume}{25}},
  \bibinfo{pages}{L201} (\bibinfo{year}{1992}), \eprint{hep-th/9111026}.

\bibitem[{\citenamefont{Moloney and Pruessner}(2003)}]{MoloneyPruessner:2003}
\bibinfo{author}{\bibfnamefont{N.~R.} \bibnamefont{Moloney}} \bibnamefont{and}
  \bibinfo{author}{\bibfnamefont{G.}~\bibnamefont{Pruessner}},
  \bibinfo{journal}{Phys.~Rev.~E} \textbf{\bibinfo{volume}{67}},
  \bibinfo{pages}{037701} (\bibinfo{year}{2003}), \eprint{cond-mat/0211240}.

\end{thebibliography}

\end{document}